\documentclass[conference]{IEEEtran}
\usepackage{float}
\usepackage{graphicx}
\usepackage{tikz}
\usetikzlibrary{positioning}
\usepackage{array}
\usepackage{enumitem}
\usepackage{amsmath,amssymb}
\usepackage[square,numbers,sort&compress]{natbib}
\usepackage{float}
\usepackage[hyphens]{url}
\usepackage{wasysym}
\usepackage{booktabs}
\usepackage{makecell}
\usepackage{tikz}
\usepackage{forest}
\usetikzlibrary{trees,positioning,shapes,shadows,arrows.meta}

\begin{document}

\title{Bitcoin Cross-Chain Bridge: A Taxonomy and\\Its Promise in Artificial Intelligence of Things}


\author{\IEEEauthorblockN{Guojun Tang}
\IEEEauthorblockA{\textit{University of Calgary} \\
Calgary, Canada \\
guojun.tang@ucalgary.ca}
\and
\IEEEauthorblockN{Carylyne Chan}
\IEEEauthorblockA{\textit{BlockSpaceForce Research	} \\
Singapore, Singapore \\
carylyne@blockspaceforce.com}
\and
\IEEEauthorblockN{Ning Nan}
\IEEEauthorblockA{\textit{BlockSpaceForce Research	} \\
Singapore, Singapore \\
nan@blockspaceforce.com}
\and
\IEEEauthorblockN{Spencer Yang}
\IEEEauthorblockA{\textit{BlockSpaceForce Research	} \\
Singapore, Singapore \\
spencer@blockspaceforce.com}
\and
\IEEEauthorblockN{Jiayu Zhou}
\IEEEauthorblockA{\textit{University of Michigan} \\
Ann Arbor, United States \\
jiayuz@umich.edu}
\and
\IEEEauthorblockN{Henry Leung}
\IEEEauthorblockA{\textit{University of Calgary} \\
Calgary, Canada \\
leungh@ucalgary.ca}
\and
\IEEEauthorblockN{Mohammad Mamun}
\IEEEauthorblockA{\textit{National Research Council Canada	} \\
Fredericton, Canada \\
mohammad.mamun@nrc-cnrc.gc.ca}
\and
\IEEEauthorblockN{Steve Drew}
\IEEEauthorblockA{\textit{University of Calgary} \\
Calgary, Canada \\
steve.drew@ucalgary.ca	}
}

\maketitle

\begin{abstract}
Bitcoin’s limited programmability constrains its integration with DeFi and multi-chain applications. This paper presents a comprehensive taxonomy of Bitcoin cross-chain bridge protocols, systematically analyzing their trust assumptions, performance characteristics, and applicability to the Artificial Intelligence of Things (AIoT) scenarios. We categorize bridge designs into three main types: naive token swapping, pegged-asset bridges, and arbitrary-message bridges. Each category is evaluated across key metrics such as trust model, latency, capital efficiency, and DeFi composability. Emerging innovations like BitVM and recursive sidechains are highlighted for their potential to enable secure, scalable, and programmable Bitcoin interoperability. Furthermore, we explore practical use cases of cross-chain bridges in AIoT applications, including decentralized energy trading, healthcare data integration, and supply chain automation. This taxonomy provides a foundational framework for researchers and practitioners seeking to design secure and efficient cross-chain infrastructures in AIoT systems.
\end{abstract}


\begin{figure*}[th]
    \centering
    \includegraphics[scale=0.55,trim={5 60 20 90},clip]{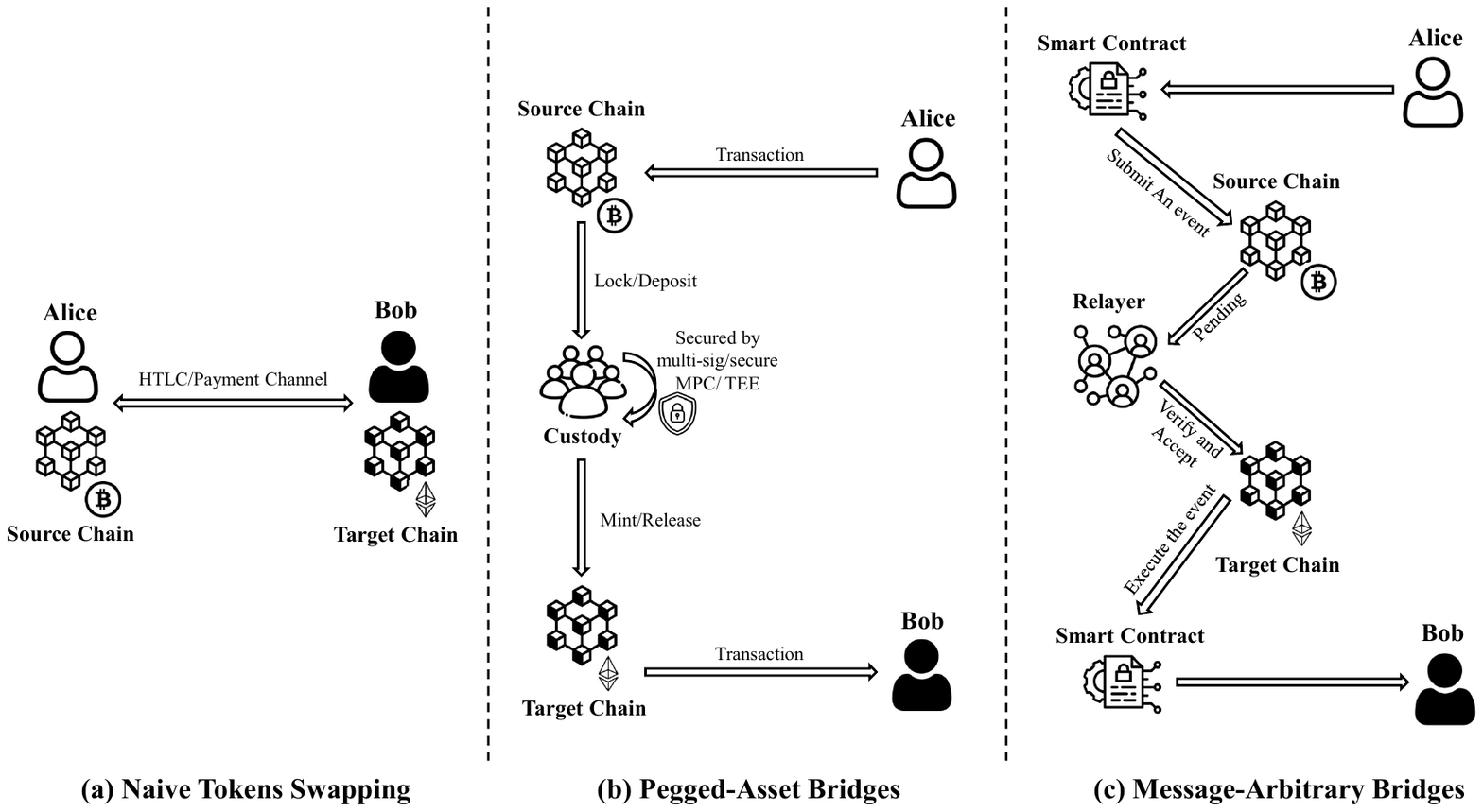}
    \caption{Overview of (a) Naive Tokens Swapping, (b) Pegged-asset Bridges, and (c) the Arbitrary-Message Bridges}
    \label{fig:overview}
\end{figure*}

\section{Introduction}\label{sec:intro}
Bitcoin \cite{bitcoin}, the first decentralized cryptocurrency,  is the symbolic leader of the crypto movement and remains the most valuable cryptocurrency. However, the Bitcoin network has limitations in its scripting capability since its inception. Its isolated network state poses challenges to direct interoperability with other blockchains. The ability to securely exchange digital assets between the Bitcoin networks onto other blockchains swiftly is increasingly desirable for enabling decentralized finance (DeFi), cross-chain asset uses, and broader blockchain ecosystem integration.

Cross-chain bridges seek to fulfill this role by enabling assets to flow securely among several widely adopted blockchain networks.
Designing a cross-chain bridge faces multi-dimensional challenges. Ensuring both \textit{low latency} and \emph{high security} is difficult. In addition, other significant properties are noteworthy in the cryptocurrency scenario, such as \emph{capital efficiency}, \emph{universality}, \emph{composability to DeFi}, and \emph{computational overhead} in the smart contract.

A fundamental obstacle is that blockchains lack a native trust mechanism between them. The classic result by Zamyatin et al. \cite{XCLAIM} shows that trustless cross-chain communication is impossible without additional assumptions, since there is no common authority to enforce atomicity across independent ledgers. Hence, all practical interoperability solutions introduce some trusted component or assumption to coordinate actions on different chains. Current approaches span a spectrum of trust models. One classic method is the \textit{hash time-lock contract} (HTLC) \cite{htlc}, which enables \textit{atomic swaps} \cite{atomic_swap,atomiccrosschainswaps}. It empowers the asset exchange between two different blockchains. HTLC is trustless, where the security relies on the blockchain consensus itself instead of the need for other third parties, but only facilitates one-time exchanges between two parties and results in substantial waiting times due to blockchain confirmation requirements. Moreover, HTLC demonstrates financial limitations, characterized by low capital efficiency due to the locking of the coins for the duration of the swapping process. Furthermore, it lacks composability with DeFi as its functionality is restricted to the mere exchange of cryptocurrencies.

The \emph{payment-channel}-based cross-chain bridge solutions use payment channels to handle cross-chain transactions without the need for a trusted third party or the blockchain, thereby accelerating cross-chain transactions. \citet{universal_swap,virtual_payment_channel} implemented the asset atomic swap via the payment channel, significantly improving the efficiency compared to the original HTLC-based method. The cross-chain nature of payment channels makes these solutions largely universal to bridge various blockchains. However, the capital efficiency is usually restricted by the hop of the route and shows poor interoperability on DeFi.

\emph{Collateralized peg protocols} like XCLAIM \cite{XCLAIM} use vault systems to lock native assets under economic security. Users receive minted tokens on a target chain backed by the locked collateral, reducing the need to trust a single custodian by distributing trust among economically incentivized participants, but introducing complex failure recovery mechanisms when liquidating the vault collateral. Compared to the HTLC-based atomic swap, these protocols are faster, but they still take hours to a day to process the asset deposit and redemption. 

Another bridging method is using \emph{on-chain relays} to transmit block information from the source chain to a targeted chain. The target chain then executes the message or mints the wrapped asset from the relay node once it verifies and accepts the relay information \cite{btcrelay, flyclient, NiPoPoW, zkbridge}. These approaches offer several benefits, including being trustless, secure, universal, and highly scalable. Yet, they still face challenges such as heavy proving costs and latency, and substantial resource consumption with high transaction fees in smart contracts.

An emerging direction is to leverage Bitcoin's scripting via new constructions to achieve cross-chain interoperability. For instance, \emph{BitVM} \cite{bitvm} and the follow-up work \emph{BitVM2} \cite{bitvm2bridge} show that one can encode arbitrary computations and even bridging logic in Bitcoin through an off-chain protocol with on-chain fraud proofs. This approach requires no changes to Bitcoin consensus and, in theory, can enable a trust-minimized bridge where Bitcoin funds are locked by a script that enforces rules via a dispute mechanism. In such a design, the safety of the bridge no longer relies on an honest majority of trustees but rather on at least one honest verifier to detect fraud. While promising, BitVM-based designs currently suffer from high on-chain overhead (multiple transactions in case of disputes) and longer worst-case latency due to interactive verification rounds.

Some alternative bridges leverage the \emph{external verifier}, which outsources verification of a source‑chain event to an off‑chain committee or service as a proxy. This proxy then produces an attestation that the destination chain’s contract checks. Axelar Network \cite{axelar} and deBridge \cite{debridge} utilize the external validator to attest the message from the source chain, while LayerZero \cite{layerzero} establishes this process through the agreement of two independent groups, oracle and relayer. They avoid custodial multisigs and per-user collateral, but do not verify foreign consensus on-chain, unlike SPV/zk light-clients \cite{zkbridge}.  

\emph{Federated sidechains} is another type of bridge controlling funds on the source chain. For example, in the Liquid Bitcoin sidechain \cite{liquid_multisig}, BTC is locked on Bitcoin in a federation-controlled 11-of-15 threshold multisig for the peg-in and peg-out process. Recently, Fractal Network \cite{fractal}, a merge-mined Bitcoin sidechain designed to scale BTC with faster blocks, higher throughput, and smart‑contract-style programmability, also showed its potential to implement the cross-chain bridge. The Fractal Network is a recursive network in which the sub‑chains can be spun up under the main chain, inheriting security and bridge infrastructure to scale horizontally. The current implementations, such as SimpleBridge \cite{simplebridge} and BoolNetwork \cite{boolbridge}, offer the bidirectional bridge from BTC to other cryptocurrencies (sBTC, bBTC, etc.). The sidechain-based method offers a clear, simple, and auditable cross-chain bridge solution with a fast peg-out latency. It also establishes good scalability and compatibility with the DeFi applications. However, it suffers from the slow peg-in latency and the trust assumption that depends on an honest supermajority of the known entities in the multisig. 

Blockchain technology, as a decentralized ledger, has been widely used in various decentralized AIoT applications, such as smart healthcare \cite{healthcare_1,healthcare_2,blockchain_integration_healthcare}, supply chains \cite{supply_chain_1, supply_chain_2,supply_chain_3,supply_chain_4}, and energy trading scenarios \cite{energy_trading_1,energy_trading_2,energy_trading_3}. Depending on the specific entities and solutions involved, various blockchain ledgers may be implemented. Ensuring interoperability among these heterogeneous ledgers presents a considerable challenge to AIoT applications. The cross-chain bridges technology provides a feasible connection among those distributed ledgers.

In this paper, we survey the current cross-chain bridges, including the methods based on HTLC atomic swap, payment channel, collateralized pegs, on-chain relay, BitVM, external verifier, and sidechains. Existing cross-chain surveys aim at different aspects. For instance, \citet{security_survey1,security_survey2} concentrated on a comprehensive analysis of the security of cross-chain bridges, whereas \citet{interoperability_survey1} emphasized a structural model and a layered architecture for cross-chain solutions. Meanwhile, \citet{interoperability_survey2, interoperability_survey3} highlighted the extensive range of literature, offering a synthesis between academia and industry derived from a substantial corpus. We then conduct an analysis of existing bridges by examining their methodologies and evaluating their pros and cons based on properties such as trust model, latency, and universality. Furthermore, the financial aspects, including capital efficiency, on-chain consumption, and the composability of DeFi, are discussed. Initially, we provide a generic form of cross-chain bridges. Subsequently, based on the mechanisms by which they exchange the asset from different blockchains, we demonstrate an innovative taxonomy for the aforementioned bridges. Following this, a literature review pertaining to each bridge category is presented, and their respective advantages and disadvantages are concluded through a comparative table. Finally, we engage in a discussion on potential cases of integrating cross-chain bridges into practical AIoT applications.


\section{Background}
In this section, we provide an introductory overview of the problem addressed by cross-chain bridges with their common characteristics. A taxonomy is then presented according to the underlying mechanism of these bridges.

\subsection{Overview of Cross-Chain Bridges}
\begin{figure}
    \centering
    \resizebox{0.48\textwidth}{!}{
\tikzset{
    basic/.style  = {draw, text width=2cm, align=center, font=\sffamily, rectangle},
    root/.style   = {basic, text width=3cm, rounded corners=2pt, thin, align=center, fill=green!30},
    onode/.style = {basic, text width=3cm, thin, rounded corners=2pt, align=center, fill=green!60,text width=3cm,},
    tnode/.style = {basic, thin, text width=3cm, align=left, fill=pink!60, align=center},
    xnode/.style = {basic, thin, text width=3cm,rounded corners=2pt, align=center, fill=blue!20},
    wnode/.style = {basic, thin, text width=3cm, align=left, fill=pink!10!blue!80!red!10, text width=6.5em},
    edge from parent/.style={draw=black, edge from parent fork right}
}

\begin{forest} for tree={
    grow=east,
    growth parent anchor=west,
    parent anchor=east,
    child anchor=west,
    edge path={\noexpand\path[\forestoption{edge},->, >={latex}] 
         (!u.parent anchor) -- +(10pt,0pt) |-  (.child anchor) 
         \forestoption{edge label};}
}
[Cross-Chain Bridges, basic,  l sep=10mm,
    [Naive Token Swapping, xnode,  l sep=10mm,
        [Atomic Swaps (HTLC), tnode,l sep=10mm]
        [Payment Channel, tnode] ]
    [Pegged-Asset Bridges, xnode,  l sep=10mm,
        [Federated Sidechains, tnode]
        [Collateralized Vault, tnode]
        [BitVM, tnode] ]
    [Arbitrary-Message Bridges, xnode,  l sep=10mm,
        [On-chain Relays, tnode]
        [External Verifier, tnode]
         ] ]
\end{forest}
}
    \caption{Taxonomy of Cross-Chain Bridges}
    \label{fig:taxonomy}
\end{figure}
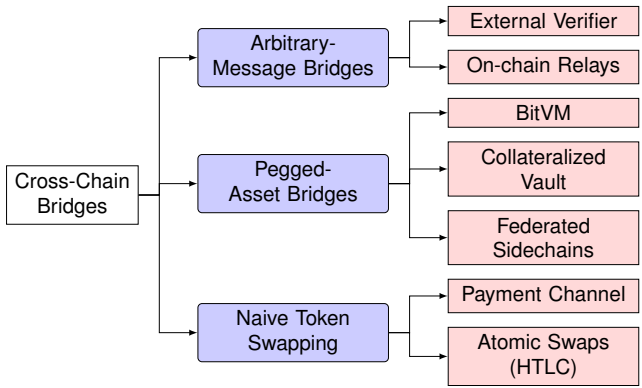

The cross-chain bridge is used to facilitate the exchange of assets between two different blockchains. Specifically, a user Alice from a source chain wants to swap her tokens/assets to Bob, who is the user from the target chain. Typically, the bridge will guarantee the properties as follows:
\begin{itemize}
    \item \textbf{Safety}: No forged messages, mints, or transfers on the target chain without a real event on the source chain.
    \item\textbf{Liveness}: If the event is final on the source chain, the target chain will execute the event.
    \item \textbf{Consistency}: No double-mints or double-releases.
\end{itemize}
Cross-chain bridges implementations with the guarantees above follow this taxonomy: \textbf{naive tokens swapping}, \textbf{pegged-asset bridges}, and the \textbf{arbitrary-message bridges}. The naive token swapping method will allow the user to directly swap their token via the coordination of HTLC smart contracts. Methods that rely on pegged-asset bridges use a relayer to manage deposits and withdrawals of cross-chain tokens. Message-arbitrary bridges provide a generic message passing protocol, which enables the source chain to transmit a message or event to the target chain, and it will be executed once it has undergone a process of approval or verification. This protocol can complete the exchange of the tokens or assets. The taxonomy of cross-chain bridges is presented in Figure \ref{fig:taxonomy} and the overview of each mechanism is illustrated in Figure \ref{fig:overview}.

\subsection{Naive Tokens Swapping}
Approaches such as HTLC-based atomic swaps \cite{atomic_swap, atomiccrosschainswaps} and payment channel-based exchanges \cite{virtual_payment_channel, universal_swap} are identified as naive token swapping, where the user will directly swap or transfer their tokens rather than a wrapped asset. Alice and Bob will first lock their own tokens on their blockchain. Once they reach an agreement, they swap their tokens atomically. This process is facilitated through the utilization of either the HTLC smart contract or the payment channel. However, because it only supports the naive tokens swapping instead of other wrapped assets, they exhibit limited composability within the context of decentralized DeFi.

\subsection{Pegged-Asset Bridges}
The concepts of collateralized vault \cite{XCLAIM}, BitVMs \cite{bitvm,bitvm2bridge}, and sidechains \cite{liquid_multisig,simplebridge,boolbridge} are categorized as forms of pegged-asset bridges. The processes of peg-in (transferring assets from the source chain to the target chain) and peg-out (redeeming assets from the target chain to the source chain) are dependent on custodial mechanisms. Users are required to deposit or lock their assets on the source chain, which subsequently allows for the release or minting of corresponding assets on the target chain. The custody is responsible for safekeeping assets and usually consists of a custody address or vault, which can be managed by cryptographic methods such as multi-signature, secure multi-party computation (MPC), or trusted execution environment (TEE).

\subsection{Message-Arbitrary Bridges}
Message-arbitrary bridges provide a generic message-passing protocol, enabling the source blockchain to send messages or events to the target blockchain, where they are executed following a verification or approval process. This protocol can facilitate the transfer of tokens or assets effectively. The typical process is as follows: Alice submits an event from the source chain. Once the event is verified, it will be executed on the target chain. This protocol can coordinate the mint or asset transfer. Methods employing external verifiers \cite{axelar,debridge, layerzero} involve verification by external verifiers, such as validators from the external blockchain or a decentralized validation network. In contrast, the verification through on-chain relays \cite{btcrelay,flyclient,NiPoPoW,zkbridge} is predominantly conducted on-chain. Due to these methods utilizing a generic message-passing protocol capable of supporting arbitrary message transmission across diverse blockchains, they exhibit significant composability with Decentralized Finance (DeFi). Furthermore, users generally do not need long-lived collateral or pre-funded channels, thereby demonstrating notable capital efficiency.

\section{Literature Review}\label{sec:related}
Blockchain interoperability has been studied extensively in both academic research and industry deployments. We summarize the most relevant approaches and technologies that influence the design of cross-chain bridges and contrast their properties, followed by a discussion of the trust model.

\subsection{Atomic Swaps and HTLCs}
\emph{Atomic cross-chain swaps} \cite{atomic_swap,atomiccrosschainswaps} allow two parties to exchange assets across different blockchains without trusting a third party. Typically, this process is coordinated via HTLCs \cite{htlc}. One party (Alice) locks coins on Chain A in a contract that can be claimed by the counterparty (Bob) if and only if Bob reveals a secret (a preimage) within a time limit; Bob similarly locks his coins on Chain B under the hash of the same secret. If either fails to receive the other's asset in time, they can reclaim their own. This protocol guarantees that either both transfers occur or neither occurs. Atomic swaps are trustless and were successfully demonstrated between Bitcoin and Ethereum \cite{btc_eth_swap}. However, they have significant limitations. First, they do not create a persistent ``bridged'' asset that can circulate on another chain. Each new user who wants to move BTC to another chain via atomic swap must find a counterparty and execute a new swap. This is impractical for scaling to many users or continuous DeFi use. Second, the operation can be slow and expensive: users generally wait for multiple block confirmations on each chain to ensure the HTLC locks are secure and to avoid fraud by double-spending before the timeout. For Bitcoin, waiting for up to 6 confirmations is common for safety, which adds significant latency.


\subsection{Payment Channels}

Payment channel, or state channel, networks \cite{state_channel} facilitate fast off-chain transfers by locking funds into multi-signature contracts and only committing the net results on-chain. \citet{universal_swap} improved the atomic swap by replacing the hash‑timelocks with cryptographic “timed” or adaptor‑style signatures. The blockchain only sees a normal signature; the atomicity is enforced off‑chain by the cryptographic relation between two signatures/secrets. \citet{virtual_payment_channel} defined a virtual payment channel that spans two heterogeneous blockchains via an intermediate node to cut off the on-chain overhead. The payment channel achieves millisecond-level transaction latency and high throughput by avoiding global consensus for each payment. Most computations are performed off-chain, which will not introduce heavy on-chain computation. However, using this network as a general bridge is limited by liquidity and the requirement of live participants. A user must find a route with sufficient capacity connecting the two networks, and intermediate nodes must be online and trustlessly incentivized to forward. Moreover, it does not create a token on the other chain; it just routes payments. For instance, if many users want to hold a Bitcoin-backed asset on, say, Ethereum via Lightning, it would require a very well-capitalized network of channels connecting Bitcoin and Ethereum, which currently does not exist. In practice, Lightning is better suited to fast payments than to locking large amounts of value long-term on another chain.


\subsection{Collateralizated Vault}\label{sec:collateralization}
Protocols like XCLAIM \cite{XCLAIM} aim to remove the need to trust a fixed federation by introducing \emph{collateralized agents} (vaults) and \emph{smart contracts} on the target chain. In XCLAIM, anyone can become a vault by locking collateral in the target chain's asset in a contract; users who want to bridge BTC send it to a Bitcoin address provided by a vault, and the contract mints an equivalent token for the user. The vault must later send back BTC when the user redeems ETH. If the vault fails, its collateral is slashed and used to compensate the user. This design ensures that as long as the vault's collateral value exceeds the BTC value, the user is economically safe. XCLAIM thus achieves trustlessness in the sense that vaults are not trusted, but introduces new assumptions that vaults are economically rational to behave honestly. Also, the security requires price feeds and over-collateralization. It also requires an active contract on the target chain to manage logic, meaning it is not purely a two-chain protocol but involves a third intermediary environment for enforcement.


\subsection{On-Chain Relays}

Another line of work is using one blockchain to verify events on another via Simplified Payment Verification (SPV) proofs or more advanced cryptographic proofs. An example was BTCRelay \cite{btcrelay}, an Ethereum contract that stored Bitcoin block headers and allowed users to submit Merkle proofs of Bitcoin transactions. With BTCRelay, any Ethereum contract could check that a given BTC was sent to a certain address, enabling trustless bridging where Ethereum acts on BTC events. However, BTCRelay had issues: the cost of continuously submitting Bitcoin block headers became prohibitive for volunteers to maintain, and there was no direct incentive for relaying headers once the novelty wore off. Additionally, BTCRelay required about 2400 gas per header byte, making it very expensive; verifying a single Bitcoin transaction inclusion could cost tens of dollars in gas. As a result, BTCRelay fell out of use. Scheme like zkBridge \cite{zkbridge} employs the light client to verify the event: the destination chain verifies a ZK proof (zk-SNARK \cite{zk_snark}) of the source chain’s headers and inclusion. Newer approaches use succinct proofs to compress chain states. FlyClient \cite{flyclient} and NiPoPoW (Non-interactive Proofs of Proof-of-Work) \cite{NiPoPoW} propose ways to prove a transaction is in the longest chain with much less data than all headers. 


\subsection{BitVM}

BitVM \cite{bitvm} is an approach that doesn't require changing Bitcoin but uses clever off-chain protocols to achieve similar outcomes. BitVM's recent iteration, BitVM2 \cite{bitvm2bridge}, specifically demonstrates a Bitcoin bridge to a hypothetical rollup by using a verification game: a set of off-chain operators maintains a sidechain, and any invalid state transition can be proven on Bitcoin via a series of transactions. The BitVM Bridge design effectively replaces the honest majority assumption on signers with an honest minority, also known as \emph{trust-minimizing}, assumption on challengers: as long as at least one watcher is honest and online to dispute fraudulent sidechain updates, BTC cannot be stolen. This guarantee is powerful, but it may lead to increased on-chain activity when disputes arise, which may result in lengthy challenge periods. BitVM2 intends to settle disputes with just a few Bitcoin transactions, contrasting with previous versions that required many. Moreover, current BitVM prototypes cannot yet manage many users or high-frequency transfers easily, since each deposit may entail setting up new off-chain contracts and potential on-chain claims.


\subsection{External Verifier}
External verifiers establish a general message passing mechanism for two different blockchains. Typically, an event on the source chain will be verified by an external verifier, which can be a blockchain, a middleware, or a decentralized validator network. Once the verifier authorizes the event, it will be submitted to the target chain and executed. Axelar Network \cite{axelar} is a proof-of-stack (PoS) blockchain that applies threshold cryptography (e.g., secure multiple parties computation) to implement the verification. Similarly, deBridge \cite{debridge}, a decentralized validator network, uses a rotating validator set to observe the source chain and collectively signs a message off‑chain to verify the event. LayerZero is a middleware protocol that chooses multiple Decentralized Verification Networks (DVNs) and requires N‑of‑M approvals to complete the verification of the event. The security model of the external verifier-based method usually relies on the validator quorum. More precisely, a proportion of validators is honest. The latency is moderately fast but still limited by validators’ observation and destination inclusion. Because those methods offer a general message passing mechanism and allow arbitrary messages, they can plug directly into DeFi. The off-chain node partially does the verification. It only requires a few on-chain computations. The general message passing mechanism only acts as a flexible intermediary to connect two different blockchains, leading to a high degree of universality.

\subsection{Sidechains}

Federated peg systems involve a fixed set of entities jointly managing assets. In Blockstream's Liquid sidechain \cite{liquid_multisig}, Bitcoin is pegged into the sidechain through a federation of functionaries: a quorum of 11 out of 15 functionaries must sign a Bitcoin transaction to move funds out of the sidechain (peg-out). To move BTC into Liquid, users send BTC to a multi-sig address controlled by the federation, and an equal amount of L-BTC is issued on Liquid. Liquid achieves fast block times and confidential transactions on the sidechain, but users must trust that the functionaries will not collude to steal the locked Bitcoin. Other examples of using sidechains include the fractal network-based methods, such as SimpleBridge \cite{simplebridge} and BoolBridge \cite{boolbridge}. The fractal network offers a merge-mined Bitcoin sidechain with faster blocks and a recursive multi-layer architecture that reuses Bitcoin core code, anchors data back to Bitcoin, and targets faster, cheaper transactions and smart-contract-style features than L1 can offer. SimpleBridge is a lightweight bridge supported by UniSat. In SimpleBridge, users lock BTC or other supported assets on Bitcoin. After sufficient confirmations, a UniSat‑operated signer set mints 1:1 representations (e.g., sBTC) on Fractal. Burning on Fractal triggers a signed Bitcoin release. Fees are usually fixed, e.g., 0.00003 BTC deposit, 0.25 FB withdrawal. Bool Network relies on the  Dynamic Hidden Committee (DHC), which is a rotating, partially hidden validator set using MPC/TEE + ZK techniques to co‑sign the peg wallet. Users bridge BTC and other cryptocurrencies, such as bBTC, with similar lock‑mint/burn‑release semantics. These sidechain-based methods show effectiveness in the peg-out process and good composability with DeFi, and their security still depends on the honest majority of the committee/federation. 
The capital efficiency is usually tailored to one L1-sidechain pair.


Table~\ref{tab:comparison} provides a high-level comparison of various cross-chain bridging approaches and their properties. We compare bridges from different dimensions, including trust model of the approach, the latency of transferring the cryptocurrencies, on-chain computational consumption (e.g., gas in Ethereum), capital efficiency, and the compatibility with DeFi. $\Circle$, $\LEFTcircle$, and $\CIRCLE$ denote the high level, medium level, and low level, respectively.

\begin{table*}[t]
\caption{Comparison of Bitcoin Cross-Chain Bridge Approaches}
\label{tab:comparison}
\centering
\begin{tabular}{lcccccc}
\toprule
\textbf{Approach} & \textbf{Trust Model} & \textbf{Latency} & \textbf{\makecell{Computation \\ Consumption}} & \textbf{Universality} & \textbf{\makecell{Capital \\ Efficiency}} & \textbf{DeFi-Compat} \\
\toprule
Atomic Swap \cite{atomic_swap,atomiccrosschainswaps} & HTLC & $\CIRCLE$ & $\Circle$  & $\Circle$ & $\Circle$ & $\Circle$  \\
Payment Channel \cite{universal_swap,virtual_payment_channel} & HTLC & $\Circle$ & $\Circle$ & $\CIRCLE$  & $\Circle$ & $\Circle$ \\
Collateralizated Vault \cite{XCLAIM} & Economic &$\LEFTcircle$ & $\LEFTcircle$ &  $\LEFTcircle$ & $\CIRCLE$ & $\CIRCLE$ \\
On-Chain Relays \cite{btcrelay,flyclient,NiPoPoW,zkbridge} & Trustless & $\Circle$ &  $\CIRCLE$ & $\CIRCLE$  &  $\CIRCLE$ &  $\CIRCLE$ \\
BitVM Bridge \cite{bitvm,bitvm2bridge} & Trust-Minimizing & $\Circle$ &  $\LEFTcircle$ &  $\CIRCLE$ & $\CIRCLE$ & $\CIRCLE$ \\
Sidechain \cite{liquid_multisig,simplebridge,boolbridge} & Majority Honest & $\LEFTcircle$ & $\LEFTcircle$ &$\LEFTcircle$ & $\CIRCLE$ & $\CIRCLE$  \\
External Verifier \cite{axelar,debridge, layerzero} & Validator Quorum & $\LEFTcircle$ & $\Circle$ & $\CIRCLE$ & $\CIRCLE$ & $\CIRCLE$  \\
\bottomrule
\end{tabular}
\end{table*}

\subsection{Trust Model of Bridges}
It is instructive to note where trust remains. Therefore, we will discuss the trust model of each bridge.

Atomic swap and payment channels rely on the security of HTLC, which provides strong security purely based on the cryptographic method, but it still suffers from the liveness issue. For example, if the counterparty disappears, the swap fails, but the user still gets a refund eventually, which leads to a temporary liquidity lock-up until the timeout. Collateralized vault like XCLAIM depends on the economic rationality that the vault has no incentive to misbehave, so it will act honestly. The security will downgrade if the vault is under-collateralized. 

Sidechain and external-verifier-based bridges guarantee security by the honest majority in the verification group. It introduces an extra security assumption to maintain the safety of the system. While the security of BitVM is trust-minimizing, unlike the honest majority, it only requires existential honesty in the verification phase. It has a weaker assumption but still relies on an external security assumption. Compared to the majority, honest and trust-minimizing, on-chain relays are trustless, where the security does not rely on any extra entity and can be reduced to the security of the source/target chain itself. It is easy to achieve the trustless security assumption. Usually, because a series of sound cryptographic methods preserves the blockchain security, it may provide a very sound security.

\section{Bridging the Cross Blockchains in AIoT Applications}

In this section, we will discuss the potential case of appropriate cross-chain bridges for specific AIoT applications, including energy trading, smart healthcare, and supply chain management. In the context of energy trading, the market necessitates a bridging mechanism that facilitates delivery-versus-payment transactions between prosumers and microgrids, periodic settlements, and bilateral transactions. Consequently, the feasibility of employing an atomic swap bridge as a cross-chain solution is demonstrated. For the healthcare case, there are various workflows that depend on distinct blockchains, such as those managing patient records, prescription records, and insurance information. An arbitrary-message bridge is deemed suitable for coordinating workflows among these disparate ledgers. Within the supply chain domain, the engagement of multiple service providers poses a challenge in transaction management and liquidity assurance, which can be addressed through a pegged-assets bridge, specifically via collateralized vaults.

\subsection{Case in Energy Trading}
In energy AIoT systems, HTLC-based or payment channel-based atomic swaps facilitate peer-to-peer energy trading across blockchain boundaries. Imagine a local microgrid ledger where solar AIoT sensors record energy credits, and a separate public blockchain where payments are made in a cryptocurrency token. A cross-chain bridge can enable, for instance, a household with surplus solar energy, tokenized on the local ledger, to exchange it directly for cryptocurrency payments from a neighbor, thereby obviating the need for an intermediary utility. The atomic swap protocol ensures that energy credits are transferred only upon securing the Bitcoin payment, and vice versa, preserving a trust-minimized transaction framework. This mechanism holds potential to support decentralized energy marketplaces where devices autonomously engage in the trading of energy or carbon credits across different networks. 

\begin{figure}[ht]
    \centering
    \includegraphics[scale=0.45,trim={155 145 160 145},clip]{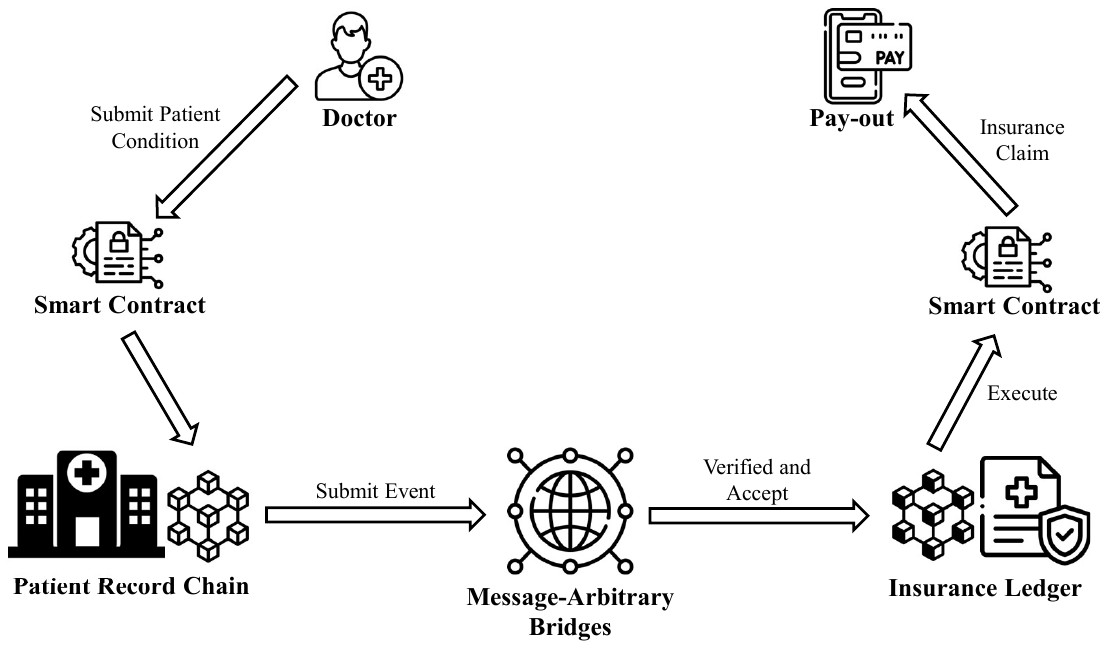}
    \caption{Potential Case of Cross-Chain Bridge in Smart Healthcare}
    \label{fig:healthcare_case}
\end{figure}

\subsection{Case in Healthcare}
In the domain of healthcare, message-arbitrary bridges can provide a message passing protocol across different blockchains, verifying credentials or events across these networks, thereby enhancing data integrity and automation. Consider a scenario involving a federated healthcare blockchain utilized by hospitals to record patient consent or device data, and a public insurance blockchain that manages claims processing. Through the deployment of this cross-chain bridge, a smart contract on the insurance blockchain can autonomously validate proof of a patient's consent transaction recorded on the hospital blockchain prior to the approval of a claim payout. This mechanism obviates the necessity of relying on the hospital's API or a centralized integration system. For example, if a patient's AIoT medical device records an emergency event on a local chain, the bridge can enable a remote telemedicine service on a separate chain to authenticate the occurrence of the event and potentially initiate an alert or payment, independent of a singular oracle. This methodology is congruent with healthcare’s need for strict data integrity and auditability when sharing records among providers \cite{blockchain_integration_healthcare}.
The potential case overview is illustrated in Figure \ref{fig:healthcare_case}.

\subsection{Case in Supply Chain}
Global supply chains often involve multiple platforms, such as a shipping AIoT platform, a trade finance blockchain, or a customs paperwork ledger \cite{supply_chain_1, supply_chain_2}. The neutrality and liquidity of cryptocurrencies, such as ETH, make them attractive as a settlement currency to bridge these systems. Collateralized vaults enable supply chain participants to utilize ETH for trade transactions across networks. For example, an exporter could lock its ETH within a vault to mint a token on a supply-chain-specific blockchain that monitors shipments. This token can be used within that network's smart contracts—potentially serving as escrow within an AIoT-supervised letter-of-credit contract. Upon the shipment’s arrival and subsequent confirmation of delivery by AIoT oracles, the escrow is released to the exporter. Due to the token being pegged, the exporter can later redeem it for actual ETH, thereby obtaining global liquidity. Meanwhile, the importer is able to transact in ETH without directly interacting with the Ethereum main chain during the procedure. Collateralized pegs give supply chain platforms a decentralized and robust common currency, resolving interoperability challenges between financial and physical flows. However, the trade-off, as noted in Section \ref{sec:collateralization}, involves the management of vault infrastructure and the requirement for over-collateralization, which introduces added complexity and expense. Participants must evaluate these factors against the advantage of avoiding dependence on a single central escrow or foreign exchange intermediary for cross-network transactions.

\section{Conclusion}
This paper presented a comprehensive taxonomy of Bitcoin cross-chain bridge protocols, highlighting the tradeoffs between trust minimization and usability. We demonstrated the practical implications of cross-chain bridging in AIoT scenarios. Future research should explore hybrid designs that combine the strengths of existing models, investigate formal security frameworks for heterogeneous trust assumptions, and develop benchmarking standards for real-world performance. Our taxonomy serves as a foundation for such investigations and a guiding framework for both researchers and practitioners in blockchain interoperability.

\bibliographystyle{unsrtnat} 
\bibliography{ref}

\end{document}